\documentclass[12pt,noshowpacs,nofootinbib,notitlepage,amsmath,superscriptaddress]{revtex4-1}
\usepackage{setspace}
\usepackage{hyperref}
\usepackage{amsfonts}
\usepackage{appendix}
\usepackage{mathtools}
\usepackage{ulem}
\usepackage{ifthen}
\newboolean{PrintVersion}
\setboolean{PrintVersion}{false}
\usepackage{amsmath,amssymb,amstext} 
\usepackage[pdftex]{graphicx} 
\usepackage[caption=false]{subfig}
\usepackage{wrapfig}
\usepackage{cleveref} 
\usepackage{enumitem}
\usepackage{float}
\usepackage{mathtools}
\usepackage{xcolor}
\usepackage[export]{adjustbox}
\usepackage{float}
\usepackage{booktabs}

\ifthenelse{\boolean{PrintVersion}}{   
\hypersetup{	
    citecolor=black,%
    filecolor=black,%
    linkcolor=black,%
    urlcolor=black}
}{} 

\let\origdoublepage\cleardoublepage
\newcommand{\clearemptydoublepage}{%
  \clearpage{\pagestyle{empty}\origdoublepage}}
\let\cleardoublepage\clearemptydoublepage

\begin{document}

\title{Charged Quark Stars and Extreme Compact Objects in Regularized 4D Einstein-Gauss-Bonnet Gravity}

\author{Michael Gammon}
\email{mgammon@uwaterloo.ca}
\affiliation{Department of Physics and Astronomy, University of Waterloo, Waterloo, Ontario, Canada, N2L 3G1}


 \author{Robert B. Mann}
\email{rbmann@uwaterloo.ca}
\affiliation{Department of Physics and Astronomy, University of Waterloo, Waterloo, Ontario, Canada, N2L 3G1}

\author{Sarah Rourke}
\email{sarah.a.rourke@mail.mcgill.ca}
\affiliation{Department of Physics, McGill University, Montreal, Quebec, Canada}

\begin{abstract}
Since the derivation of a well-defined $D\rightarrow 4$ limit for 4 dimensional Einstein Gauss-Bonnet (4DEGB) gravity  coupled to a scalar field, there has been interest in testing it as an alternative to Einstein's general theory of relativity. Using the Tolman-Oppenheimer-Volkoff (TOV) equations modified for charge and 4DEGB gravity, we model the stellar structure of charged, non-interacting quark stars. We find that increasing the Gauss-Bonnet coupling constant $\alpha$ or the charge $Q$ both tend to increase the mass-radius profiles of quark stars described by this theory, allowing a given central pressure to support larger quark stars  in general.  We also derive a generalization of the Buchdahl bound for charged stars in 4DEGB gravity.  As in the uncharged case, we find that quark stars can exist below the general relativistic Buchdahl bound (BB) and Schwarzschild radius $R=2M$, due to the lack of a mass gap between black holes and compact stars in the 4DEGB theory. Even for $\alpha$ well within current observational constraints, we find that quark star solutions in this theory can describe Extreme Compact Charged Objects (ECCOs), objects whose radii are smaller than what is allowed by general relativity.
\end{abstract}

\maketitle

\newpage

\section{Introduction}

Modified theories of gravity  continue to attract attention despite the empirical success of general relativity (GR). These theories are motivated by a variety of problems, including addressing issues   in modern cosmology \cite{bueno2016,sotiriou_2010,nojiribook,clifton_2012}, quantizing gravity \cite{ahmed2017,stelle1977}, eliminating singularities \cite{brandenberger1992nonsingular,BRANDENBERGER_1993,brandenberger1995implementing}, and, perhaps most importantly, finding viable phenomenological competitors against which GR can be tested in the most stringent manner possible.

Higher curvature theories (or HCTs) are amongst the most popular modifications. An HCT modifies the assumed linear relationship in GR between the curvature and the stress-energy, replacing the former with  an arbitrary sum of powers of the curvature tensor  (appropriately contracted to two indices). Such modifications 
provide us with a foil to further challenge the  empirical success of GR, while also making new testable predictions.

Lovelock theories \cite{lovelock1971} have long been at the forefront of this search, since they possess the distinctive feature of having 2nd order differential equations of motion. The physical significance of such theories has been unclear, however, since their higher order terms yield non-trivial contributions to the equations of motion only in more than four  spacetime dimensions  ($D>4$). 

Recently this restriction was circumvented \cite{hennigar_2020_on,Fernandes:2020nbq} in the quadratic case, or what is better known as  ``Einstein-Gauss-Bonnet" gravity .  The  Gauss-Bonnet (GB) contribution to the gravitational action is
\begin{equation}\label{eq:dgbaction}
S_{D}^{G B} = \alpha \int d^{D} x \sqrt{-g} \left[R^{\mu\nu\rho\tau}R_{\mu\nu\rho\tau} - 4 R^{\mu\nu}R_{\mu\nu} + R^2 \right]
\equiv \alpha \int d^{D} x \sqrt{-g} \mathcal{G}
\end{equation}
(where  $R_{\mu\nu\rho\tau}$ is the Riemann curvature tensor), which becomes the integral of a total derivative in $D=4$, and thus cannot  contribute to a system's gravitational dynamics in less than five dimensions. For this reason it is often referred to as a ``topological term" having no relevance to physical problems.  Indeed, the  Lovelock theorem \cite{lovelock1971} ensures that a $D=4$ dimensional metric theory of gravity must incorporate additional fields in order to have second order equations of motion and diffeomorphism invariance.  

A few years ago it was noted  \cite{glavan2020} that several exact solutions to $D$-dimensional Einstein-Gauss Bonnet gravity have a sensible limit under the rescaling 
\begin{equation}\label{eq:alpharescale}
	\lim_{D \to 4} (D-4) \alpha \rightarrow \alpha,
\end{equation}
of the Gauss-Bonnet coupling constant.  Using this approach a variety of 4-dimensional metrics can be obtained, including cosmological  \cite{glavan2020,li2020,kobayashi2020}, spherical black hole \cite{glavan2020,kumar2022,fernandes2020,kumar2020,kumar2022_2}, collapsing \cite{malafarina2020}, star-like  \cite{doneva2021,charmousis2022}, and radiating  \cite{ghosh2020} metrics, each carrying imprints of the quadratic curvature effects of  their $D > 4$ counterparts.  However a number of objections to this approach were subsequently raised \cite{gurses2020,ai2020,shu2020}, based on the fact that 
the existence of a limiting solution does not imply the existence of a well-defined 4D theory whose field equations have that solution. This shortcoming was quickly addressed  when it was shown that the $D\to 4$ limit in \eqref{eq:alpharescale} can be taken in the gravitational action \cite{hennigar_2020_on,Fernandes:2020nbq}, generalizing an earlier  procedure employed in obtaining the $D\to 2$ limit of GR \cite{Mann:1992ar}. One can also   compactify  $D$-dimensional Gauss-Bonnet gravity on a $(D-4)$-dimensional maximally symmetric space and then use \eqref{eq:alpharescale} to obtain a
$D=4$ HCT \cite{Lu:2020iav}. This approach yields the same result (up to trivial field redefinitions), in addition to terms depending on the curvature of the maximally symmetric $(D-4)$-dimensional space. Taking this to vanish yields
\begin{equation}\label{eq:4DEGBaction}
\begin{aligned}
S_{4}^{GB}
=&\alpha \int d^{4} x \sqrt{-g}\left[  \phi \mathcal{G}+4 G_{\mu \nu} \nabla^\mu \phi \nabla^\nu \phi-4(\nabla \phi)^2 \square \phi+2(\nabla \phi)^4\right]
\end{aligned}
\end{equation}
where we see that an additional scalar field $\phi$ appears. Surprisingly, the spherically symmetric black hole solutions to the field equations match those from the naïve $D \rightarrow 4$ limit of 
$D>4$ solutions \cite{glavan2020}. The resultant 4D scalar-tensor theory is a particular type of Horndeski theory \cite{horndeski1974}, and solutions to its equations of motion can be obtained without ever referencing a higher dimensional spacetime \cite{hennigar_2020_on}. 

We are interested here in what is called 4D Einstein-Gauss-Bonnet gravity (4DEGB), whose action is given by \eqref{eq:4DEGBaction} plus the Einstein-Hilbert term:
\begin{align}
S&= S^{GR} + S_{4}^{GB}
\nonumber\\
&= 
\frac{1}{8\pi G}
\int d^{4} x \sqrt{-g}\left[R + 
\alpha\left\{  \phi \mathcal{G}+4 G_{\mu \nu} \nabla^\mu \phi \nabla^\nu \phi-4(\nabla \phi)^2 \square \phi+2(\nabla \phi)^4 \right\}\right]\label{4DEGB}
\end{align}
which has been shown to be   an interesting phenomenological competitor to GR \cite{Clifton:2020xhc,charmousis2022,Zanoletti:2023ori}. 
Despite much exploration \cite{Fernandes:2022zrq}, the role played by these
higher curvature terms in real gravitational dynamics is still not fully understood. One important arena for testing such theories against standard general relativity is via observations of compact astrophysical objects like neutron stars. The correct theory should be able to accurately describe recent gravitational wave observations of astrophysical objects existing in the mass gap between the heaviest compact stars and the lightest black holes. 

 Modern observational astrophysics is rich in findings of compact objects and as such our understanding of highly dense gravitational objects is rapidly advancing. However, there is as of yet no strong consensus on their underlying physics. 
 A number of such objects have been recently observed that are inconsistent with standard GR and a simple neutron star equation of state. It was recently shown \cite{zhang_2021_unified,zhang_2021_stellar} that in standard general relativity, the secondary component of the merger GW190814 could feasibly be a quark star with an interacting equation of state governed by a single parameter $\lambda$.  This parameterization of strong interaction effects was inspired by another recent theoretical study showing that non-strange quark matter (QM) could feasibly be the ground state of baryonic matter at sufficient density and temperature \cite{holdom_2018_quark}. Similar analyses with a different equation of state and/or QM phase \cite{miao_2021_bayesian,lopes_2022_onthe,oikonomou_2023_colourflavour} found similarly promising results.
This same object was subsequently shown to be well described as a slowly-rotating neutron star in the 4DEGB theory without resorting to exotic quark matter EOSs, while also demonstrating that the equilibrium sequence of neutron stars asymptotically matches the black hole limit, thus closing the mass gap between NS/black holes of the same radius \cite{charmousis2022}.  More recently, some groups have also been interested in modelling ECOs (extreme compact objects) \cite{zhang_2023_rescaling,gammon_2024,Mann:2021mnc,Conklin:2021cbc} as well as unusually light compact stars \cite{horvath_2023_alight,oikonomou_2023_colourflavour} (like that in the gamma ray remnant J1731-347, which is inconsistent with minimum mass calculations of neutron stars generated by iron cores) as quark stars to explain their unusual properties. 


 To further illuminate the range of possibilities, we consider in this paper charged quark star solutions to the 4DEGB theory,  as the confining nature of the strong interaction make quark stars one of the few remaining candidates for charged stellar objects.  
 Although charged quark star solutions have previously been considered in the context of 4DEGB \cite{pretel_2022}, the upper limit of the coupling $\alpha$ was taken to be much smaller than that allowed by current observational constraints \cite{Fernandes:2022zrq,Clifton:2020xhc}, and the relation with the charged 4DEGB Buchdahl bound was not discussed (as the bound was not derived). In considering values of $\alpha$ closer to presently allowed bounds, we obtain a number of interesting novel results. 

Our most intriguing result (extending similar results found in \cite{gammon_2024}) is that charged quark stars in 4DEGB can be Extreme Compact Charged Objects (ECCOs), objects whose radii are smaller than that allowed by the Buchdahl bound and Schwarzschild radius in GR. Indeed, there exist uncharged quark stars in 4DEGB whose radii are smaller than that of a corresponding black hole of the same mass in GR\footnote{This phenomenon was also shown to be present for neutron stars
\cite{charmousis2022}, though the relationship with the Buchdahl bound was not noted.}. Observations of these latter objects, apart from indicating a new class of astrophysical phenomena \cite{Mann:2021mnc}, would provide strong evidence for 4DEGB as a physical theory. These ECOs respect a generalization of the Buchdahl bound, whose small-radius limit is that of the horizon radius of the corresponding minimum mass black hole. ECCOs likewise respect a  generalization of the charged Buchdahl bound as we shall demonstrate.

We also find in general that for a given central pressure, charged quark stars in 4DEGB have a larger mass and radius than their GR counterparts with the same pressure.  This can be attributed to the `less attractive' nature of gravity in 4DEGB: for $\alpha>0$ the gradient of the effective potential yields a weaker force than pure GR. We consequently find a larger maximal mass for a given value of $Q$ in 4DEGB than in general relativity. Similarly, increasing $Q$ also weakens the overall force gradient compared to the uncharged case, and leads to larger quark stars relative to their uncharged counterparts.

The outline of our paper is as follows: In section \ref{sec:theory} we introduce the basic theory underlying 4DEGB gravity, as well as the non-interacting quark matter equation of state that we make use of. Following this, a electromagnetic perfect fluid stress-energy tensor is employed to derive the charged 4DEGB TOV equations, and current observational constraints on the coupling constant are briefly discussed.  We use section \ref{sec:buch} to derive a generalization of Buchdahl's bound for charged stars in 4DEGB gravity to be plotted alongside the solutions to the modified TOV equations.  Section \ref{sec:results} outlines the results of our numerical calculations and briefly compares these solutions to observational data of real candidate quark stars. We conclude our results with a brief analysis of the stability of charged 4DEGB quark stars. Section \ref{sec:summary} summarizes our key findings and suggests topics for future study.

\section{Theory}\label{sec:theory}

\subsection{4D Einstein-Gauss-Bonnet Gravity}

To model the structure of a quark star in 4DEGB gravity, we add to 
the action \eqref{4DEGB}
a perfect-fluid term corresponding to the energy of the quark matter and
\begin{equation}
   S_\mathrm{Max} = 
-\frac{1}{8\pi G}   \int \sqrt{-g} \left( {\frac{1}{4}} 
 F_{\mu\nu} F^{\mu\nu} + A_\mu J^\mu  
   \right) 
\end{equation}
where $J^\mu$ is the electromagnetic current of the charged quark matter.   
 The  field equations of 4DEGB are obtained from a straightforward variational principle applied to this action. 
 
 Variation with respect to the scalar $\phi$ yields
\begin{equation}\label{eq:eomscalar}
\begin{aligned}
\mathcal{E}_{\phi}=&-\mathcal{G}+8 G^{\mu \nu} \nabla_{\nu} \nabla_{\mu} \phi+8 R^{\mu \nu} \nabla_{\mu} \phi \nabla_{\nu} \phi-8(\square \phi)^{2}+8(\nabla \phi)^{2} \square \phi+16 \nabla^{a} \phi \nabla^{\nu} \phi \nabla_{\nu} \nabla_{\mu} \phi \\
&\qquad +8 \nabla_{\nu} \nabla_{\mu} \phi \nabla^{\nu} \nabla^{\mu} \phi \\
=& \; 0
\end{aligned}
\end{equation}
while the variation with respect to the metric gives
\begin{equation}\label{eq:eommetric}
\begin{aligned}
\mathcal{E}_{\mu \nu} &=\Lambda g_{\mu \nu}+G_{\mu \nu}+\alpha\left[\phi H_{\mu \nu}-2 R\left[\left(\nabla_{\mu} \phi\right)\left(\nabla_{\nu} \phi\right)+\nabla_{\nu} \nabla_{\mu} \phi\right]+8 R_{(\mu}^{\sigma} \nabla_{\nu)} \nabla_{\sigma} \phi+8 R_{(\mu}^{\sigma}\left(\nabla_{\nu)} \phi\right)\left(\nabla_{\sigma} \phi\right)\right.\\
&-2 G_{\mu \nu}\left[(\nabla \phi)^{2}+2 \square \phi\right]-4\left[\left(\nabla_{\mu} \phi\right)\left(\nabla_{\nu} \phi\right)+\nabla_{\nu} \nabla_{\mu} \phi\right] \square \phi-\left[g_{\mu \nu}(\nabla \phi)^{2}-4\left(\nabla_{\mu} \phi\right)\left(\nabla_{\nu} \phi\right)\right](\nabla \phi)^{2} \\
&+8\left(\nabla_{(\mu} \phi\right)\left(\nabla_{\nu)} \nabla_{\sigma} \phi\right) \nabla^{\sigma} \phi-4 g_{\mu \nu} R^{\sigma \rho}\left[\nabla_{\sigma} \nabla_{\rho} \phi+\left(\nabla_{\sigma} \phi\right)\left(\nabla_{\rho} \phi\right)\right]+2 g_{\mu \nu}(\square \phi)^{2} \\
& -4 g_{\mu \nu}\left(\nabla^{\sigma} \phi\right)\left(\nabla^{\rho} \phi\right)\left(\nabla_{\sigma} \nabla_{\rho} \phi\right)+4\left(\nabla_{\sigma} \nabla_{\nu} \phi\right)\left(\nabla^{\sigma} \nabla_{\mu} \phi\right) \\ 
&\left. -2 g_{\mu \nu}\left(\nabla_{\sigma} \nabla_{\rho} \phi\right)\left(\nabla^{\sigma} \nabla^{\rho} \phi\right)
+4 R_{\mu \nu \sigma \rho}\left[\left(\nabla^{\sigma} \phi\right)\left(\nabla^{\rho} \phi\right)+\nabla^{\rho} \nabla^{\sigma} \phi\right] \right]\\
&=\;T_{\mu \nu}
\end{aligned}
\end{equation}
where $T_{\mu \nu}$ is the stress-energy tensor of the charged quark matter and the electromagnetic field, and
\begin{equation}\label{eq:gbtensor}
    \begin{aligned}
    H_{\mu \nu}=2\Big[R R_{\mu \nu}-2 R_{\mu \alpha \nu \beta} R^{\alpha \beta}+R_{\mu \alpha \beta \sigma} R_{\nu}^{\alpha \beta \sigma}-2 R_{\mu \alpha} R_{\nu}^{\alpha} -\frac{1}{4} g_{\mu \nu}\mathcal{G}
\Big]
    \end{aligned}
\end{equation}
 is  the Gauss-Bonnet tensor. These field equations satisfy the following relationship
\begin{equation}\label{eq:fieldeqntrace}
g^{\mu \nu}T_{\mu \nu}=g^{\mu \nu} \mathcal{E}_{\mu \nu}+\frac{\alpha}{2} \mathcal{E}_{\phi}=4 \Lambda-R-\frac{\alpha}{2} \mathcal{G}
\end{equation}
which can act as a useful consistency check to see whether prior solutions generated 
via the solution-limit method \cite{glavan2020} are even possible solutions to the theory. 
For example, using \eqref{eq:fieldeqntrace}
it is easy to verify that the rotating metrics generated from a Newman-Janis algorithm \cite{kumar2020,wei2020} are not solutions to the field equations of the scalar-tensor 4DEGB theory.

\subsection{Charged 4DEGB TOV Equations}
 Throughout the following we assume a non-interacting quark equation of state
\begin{equation}\label{eq:eos}
    P(r) = \frac{1}{3}(\rho - 4 B_\mathrm{eff})
\end{equation}
where $B_\mathrm{eff}$ is the MIT bag constant (for which we use a benchmark value of $60 \; \mathrm{MeV}/\mathrm{fm}^3$ \cite{zhang_2021_stellar}).

The standard Tolman-Oppenheimer-Volkoff (TOV) equations for stellar structure are well-known in GR.  Variation of the action with respect to the gauge potential $A_\mu$, together with \eqref{eq:eomscalar} and
\eqref{eq:eommetric} yield the TOV equations for the charged 4DEGB quark star. To obtain these we begin with a static, spherically symmetric metric ansatz in natural units ($G = c = 1$):

\begin{equation}
d s^2=-e^{\Phi (r)} c^2 d t^2+ e^{\Lambda (r)} d r^2+r^2 d \Omega^2.
\end{equation}

As usual \cite{hennigar_2020_on,gammon_2024}, so long as $e^{\Phi(r)} = e^{-\Lambda(r)}$ outside the star, the combination $\mathcal{E}_0^0-\mathcal{E}_1^1$ of the field equations can be used to derive the following equation for the scalar field:
\begin{equation}\label{phivac}
    \left(\phi^{\prime 2}+\phi^{\prime \prime}\right)\left(e^{\Lambda(r)}-\left(r \phi^{\prime}-1\right)^2 \right)=0.
\end{equation}
which, apart from the irrelevant $\phi=\ln \left(\frac{r-r_0}{l}\right)$ (with $r_0$ and $l$ being constants of integration), 
has the solution
\begin{equation}\label{eq:scalar}
\phi_{ \pm}=\int  \frac{1 \pm e^{\Lambda(r)/2}}{r} d r
\end{equation}
where  $\phi_-$
falls off as as $\frac{1}{r}$ in asymptotically flat spacetimes.  Choosing 
$\phi = \phi_-$ ensures  that \eqref{eq:eomscalar} is automatically satisfied.

Modelling the charged quark matter by a perfect fluid electromagnetic matter source, the  stress-energy tensor is (setting $G=c=1$) 
\begin{equation}
T_\nu^\mu=(P+\rho) u^\mu u_\nu+P \delta_\nu^\mu+\frac{1}{4 \pi}\left(F^{\mu \alpha} F_{\alpha \nu}-\frac{1}{4} \delta_\nu^\mu F_{\alpha \beta} F^{\alpha \beta}\right),
\end{equation}
from which we obtain the equations
\begin{align}\label{eq:fett}
    &\frac{e^{-2 \Lambda(r)} \left(r^2 e^{\Lambda(r)} \left(1-e^{\Lambda(r)}\right)-r^3 e^{\Lambda(r)} \Lambda'(r)\right)}{r^4} \\
    &+\alpha\frac{  e^{-2 \Lambda(r)} \left(-2 r e^{\Lambda(r)} \Lambda'(r)+2 r \Lambda'(r)-e^{\Lambda(r)} \left(1-e^{\Lambda(r)}\right)-e^{\Lambda(r)}+1\right)}{r^4} = -\rho (r)-\frac{E(r)^2}{8\pi} \nonumber\\
    &\frac{e^{-2 \Lambda(r)} \left(r^3 e^{\Lambda(r)} \Phi'(r)+r^2 e^{\Lambda(r)} \left(1-e^{\Lambda(r)}\right)\right)}{r^4} \nonumber\\
    &+\alpha\frac{  e^{-2 \Lambda(r)} \left(2 r e^{\Lambda(r)} \Phi'(r)-2 r \Phi'(r)-e^{\Lambda(r)} \left(1-e^{\Lambda(r)}\right)-e^{\Lambda(r)}+1\right)}{r^4}=P(r)-\frac{E(r)^2}{8\pi}
    \label{eq:ferr}
\\
\label{eq:cons}
    &(\rho(r)+P(r))=-\frac{2}{\Phi'(r)}\left(\frac{\mathrm{d} P}{\mathrm{~d} r}-\frac{q}{4 \pi r^4} \frac{\mathrm{d} q}{\mathrm{~d} r}\right)
\end{align}
in 4DEGB. 

Imposing asymptotic flatness 
implies that $\Phi(\infty)=\Lambda(\infty)=0$, and regularity at the center of the star implies $\Lambda(0)=0$. Using the $tt$ field equation it is easy to show that
\begin{equation}\label{eq:metricB}
    e^{-\Lambda(r)}=1+\frac{r^2}{2 \alpha}\left[1-\sqrt{1+4\alpha\left(\frac{2 m(r)}{r^3}-\frac{q(r)^2}{r^4}\right)}\right],
\end{equation}
in agreement with what was found in \cite{pretel_2022}, recalling that $E(r)^2=\frac{q(r)^2}{r^4}$. With this we arrive at the 4DEGB modified TOV equations, namely
\begin{align}
\frac{dq}{dr} &= 4 \pi r^2 \rho_e e^\frac{\Lambda(r)}{2} \label{eq:dqdr}\\
\frac{dm}{dr} &= 4 \pi r^2 \rho(r) + \frac{q(r)}{r}\frac{dq}{dr} \label{eq:dmdr}\\
\frac{dP}{dr}  &=(P(r)+\rho (r))\frac{ \left(r^3 (\Gamma +8 \pi  \alpha  P(r)-1)-2 \alpha  m(r)\right)}{\Gamma  r^2 \left((\Gamma -1) r^2-2 \alpha \right)}+\frac{q(r) }{4 \pi  r^4}\frac{dq}{dr}\label{eq:dpdr}
\end{align}
where $\Gamma=\sqrt{1+4\alpha\left(\frac{2 m(r)}{r^3}-\frac{q(r)^2}{r^4}\right)}$ (matching those found in \cite{pretel_2022}). The vacuum solution is given by $m(r) = M$ and $q(r)=Q$
where $M$ and $Q$ are constants (the total mass and charge, respectively), implying that $\Phi = -\Lambda$. Writing $e^{- \Lambda(r)}=1+2\varphi(r)$, we can   compute the gravitational force in 4DEGB due to a spherical body
\begin{equation}\label{force}
    \vec{F} = -\frac{d\varphi}{dr}\hat{r} = -\frac{r}{2 \alpha} \left(1-\frac{2 \alpha  M+r^3}{r \sqrt{8 \alpha  r M-4 \alpha  Q^2+r^4}}\right)\hat{r}\, ,
\end{equation}
which is smaller in magnitude than its Newtonian $\alpha=0$ counterpart
($\vec{F}_N = -\frac{M}{r^2}\hat{r}$) for $\alpha>0$. The force in 
\eqref{force} vanishes at 
$r=(\alpha M)^{1/3}$ if $Q=0$, but this is always at a smaller value of $r$ than the outer horizon 
$R_h = M +\sqrt{M^2-Q^2 -\alpha}$
of the corresponding black hole. Hence the gravitational force outside of any spherical body, while weaker than that in GR, is always attractive provided $\alpha>0$.  If $\alpha<0$ then the corresponding gravitational force is more attractive than in GR.  Similarly for nonzero charge, if all other parameters are held constant, increasing the charge weakens the
gravitational attraction, which only vanishes  in regions disallowed by the black hole horizon. 

Rescaling the various quantities using
\begin{equation}\label{rs1}
\bar{\rho}=\frac{\rho}{4 B_{\mathrm{eff}}}\qquad \bar{p}=\frac{p}{4 B_{\mathrm{eff}}}
\qquad \bar{\rho}_e=\frac{\rho_e}{4 B_{\mathrm{eff}}}
\end{equation}

and
\begin{equation}\label{rs2}
\bar{m}=m \sqrt{4 B_{\mathrm{eff}}}  \qquad \bar{r}=r \sqrt{4 B_{\mathrm{eff}}} \qquad \bar{q} = q \sqrt{4 B_{\mathrm{eff}}} \qquad \bar{\alpha} = \alpha \cdot 4 B_{\mathrm{eff}},
\end{equation}
we obtain  the dimensionless equations 
\begin{align}
\frac{d\bar{q}}{d\bar{r}} &= 4 \pi \bar{r}^2 \bar{\rho}_e e^\frac{\Lambda(\bar{r})}{2} 
\label{qdiff}
\\
\label{pdiff}
\frac{d\bar{m}}{d\bar{r}} &= 4 \pi \bar{r}^2 \bar{\rho}(\bar{r}) + \frac{\bar{q}(\bar{r})}{\bar{r}}\frac{d\bar{q}}{d\bar{r}} \\
\frac{d\bar{P}}{d\bar{r}}  &=(\bar{P}(\bar{r})+\bar{\rho} (\bar{r}))\frac{ \left(\bar{r}^3 (\Gamma +8 \pi  \bar{\alpha}  \bar{P}(\bar{r})-1)-2 \bar{\alpha}  \bar{m}(\bar{r})\right)}{\Gamma  \bar{r}^2 \left((\Gamma -1) \bar{r}^2-2 \bar{\alpha} \right)}+\frac{\bar{q}(\bar{r}) }{4 \pi  \bar{r}^4}\frac{d\bar{q}}{d\bar{r}}
\label{mdiff}
\end{align}
which may be solved numerically.
In the limit $\alpha \to 0$, the above equations reduce back to the well-known TOV equations for a charged, static, spherically symmetric gravitating body in GR. \\

To solve \eqref{qdiff}, \eqref{pdiff} and \eqref{mdiff} numerically we impose the boundary conditions 
\begin{equation}\label{bcs}
m(0)=0, \quad q(0)=0, \quad \rho(0)=\rho_{\mathrm{c}},
\end{equation}
where the star's surface radius $R$ is defined via $\bar{p}(\bar{R}) = 0$, namely the radius at which the pressure goes to 0 ({\it i.e.} $p(R) = 0$). We similarly define the total mass of the star to be $M = m(R)$.  

For the charge density there are several benchmark models used in the literature. We shall restrict ourselves to the model in which  the charge density is proportional to energy density (ie. $\rho_e = \gamma \rho$ or $\bar{\rho}_e = \gamma \bar{\rho}$ where $0\leq \gamma \leq 1$) \cite{zhang_2021_stellar,ray2003,arbanil2015}. Another  popular charge model sets charge proportional to spatial volume. However, this leads to exotic pressure profiles that do not decrease monotonically \cite{zhang_2021_stellar}, and thus are not consistent with arguments we shall present in section \ref{sec:buch}. As we are primarily interested in how charged 4DEGB quark stars affect the modified Buchdahl bound (since many novel results were noticed in the uncharged case \cite{gammon_2024}), we consider only the model where  $\bar{\rho}_e = \gamma \bar{\rho}$. Furthermore, since the interaction with the Buchdahl bound is the main feature of interest, we choose to present results with a fixed $Q$ rather than a fixed $\gamma$, with this charge parameter taking on standard values from the literature ($Q=(0,1,2)\times10^{20} \mathrm{\;C}$, $\bar{Q}=(0, 1.538, 3.076)\times10^{-2}$) for comparison \cite{arbanil2015,zhang_2021_stellar}.  This is done by solving the equations for a trial value of $\gamma$ and iterating until our desired fixed charge is attained.
 
With the above, numerical solutions can thus be obtained by scanning through a range of values of $\rho_c$ and solving for the star's total mass and radius.

Before proceeding to solve the TOV equations, we consider the behaviour of the scalar field $\phi$ in the interior.  By continuity with the exterior solution, the scalar is still described by \eqref{eq:scalar} inside the star \cite{charmousis2022}. In considering the interior behaviour it is instructive to rewrite $q(r) \to r \tilde{q}(r)$ (which can be done by virtue of equation \eqref{eq:dqdr} and the fact that $\lim_{r\to 0}e^{\frac{\Lambda(r)}{2}} \sim \mathrm{const} + \mathcal{O}(r)$). Inserting the   interior solution \eqref{eq:metricB}
 into \eqref{phivac} and making the latter substitution, we find
\begin{equation}
\begin{aligned}
    \phi'(r) &= -\sqrt{\frac{m(0)}{2 \alpha r}}-\frac{3 m(0)}{4 \alpha }+\frac{m(0) \sqrt{r} \left(\alpha  \left(\tilde{q}(0)^2-2 m'(0)\right)-5 m(0)^2\right)}{4 \sqrt{2} (\alpha  m(0))^{3/2}}\\
    &+\frac{r \left(4 \alpha  \left(-6 m'(0)+3 \tilde{q}(0)^2+2\right)-35 m(0)^2\right)}{32 \alpha ^2}+\mathcal{O}\left(r^{3/2}\right)
\end{aligned}
\end{equation}
Furthermore, provided 
$m(r)$ and $q(r)$ both vanish at least quadratically in $r$ for small $r$ (which is ensured
from \eqref{eq:dqdr} and \eqref{eq:dmdr} for the boundary conditions \eqref{bcs})
we find that near the origin
\begin{equation}
\begin{aligned}
    \lim_{r\to 0}\phi'(r) &\sim - \sqrt{\frac{\mathcal{M}(0)}{2\alpha }} \sqrt{r}+\frac{r}{4 \alpha }+\mathcal{O}\left(r^{3/2}\right) \approx 0\\
    \lim_{r\to 0}\phi(r) &
    \sim 
   - \frac{r^{3/2}}{3}  \sqrt{\frac{2\mathcal{M}(0)}{\alpha }} +  \frac{r^2}{8 \alpha }  + K \approx K
\end{aligned}
\end{equation}
(where $\mathcal{M}(r)=r^{-2}m(r)$ 
and $K$ is a constant) and thus regularity of the scalar at the origin is ensured.

Finally we note that the  effective bag constant   $B_\mathrm{eff} = 60$ MeV/$\mathrm{fm}^3$  can be converted to units of inverse length squared 
(`gravitational units')
with the factor $G/c^4$, yielding
\begin{equation}
    B_\mathrm{eff} = 7.84 \times 10^{-5} \; \mathrm{km}^{-2}.
\end{equation}

\subsection{Observational Constraints on the 4DEGB Coupling Constant}

An investigation of the observational constraints on the coupling $\alpha$ yielded
\cite{Clifton:2020xhc,Fernandes:2022zrq}
\begin{equation}\label{eq:constraints}
-10^{-30}\; \textrm{m}^2 < \alpha<10^{10}\; \textrm{m}^2
\end{equation}
where   the lower bound comes from ``early universe cosmology and atomic nuclei"
data \cite{charmousis2022}, 
and the upper bound follows from LAGEOS satellite observations. Regarding the lower bound as negligibly close to zero,
the dimensionless version of \eqref{eq:constraints} reads 
\begin{equation}\label{eq:dimlessconstraints}
0<\bar{\alpha} \lesssim 3.2.
\end{equation}

We note that inclusion of preliminary calculations on recent GW data suggest these constraints could potentially tighten   to $0<\alpha \lesssim 10^7\; \textrm{m}^2$, or alternatively $0<\bar{\alpha} \lesssim 0.0032$. This would mean  that deviations from GR due to 4DEGB would only be detectable in extreme environments such as in the very early universe or near the surface of extremely massive objects.  
Even   tighter bounds were assumed in previous studies of quark stars, where only solutions with $\alpha$ below 6 $\mathrm{km}^2$ ($\bar{\alpha} \leq 0.0019$) were considered \cite{banerjee_2021_strange,banerjee_2021_quark,pretel_2022}.
Adopting such a tight bound would make  compact stars near the upper end of the mass gap an ideal candidate for investigation the effects of 4DEGB theory.  

At this point in time such tighter bounds are not warranted.  A proper study of the effects of gravitational radiation in 4DEGB has yet to be carried out.  In view of this we shall assume the bound  \eqref{eq:dimlessconstraints}, which has strong observational support \cite{Clifton:2020xhc,Fernandes:2022zrq}.

\section{Generalization of the Buchdahl Bound for Charged Compact Stars in 4DEGB Gravity}\label{sec:buch}

\subsection{Derivation of a Maximal Mass for Charged Compact Objects in 4DEGB Gravity}

A generalization of the isotropic Buchdahl bound has been derived for charged stars in GR \cite{bohmer_2007} and for uncharged stars in 4DEGB theory \cite{chakraborty_2020}. Similarly, a more general mass/radius inequality was derived in GR for both charged \cite{Andreasson_2008} and uncharged \cite{Andreasson_2008_2} spheres with non-isotropic pressure profiles, where the assumption of monotonically decreasing pressure profiles can actually be discarded. In the following we extend these calculations for the case of an isotropic charged star in 4DEGB gravity, using an approach similar to Buchdahl, assuming that the pressure profiles do decrease monotonically.

By virtue of equation \eqref{eq:metricB} it is straightforward to check that
\begin{equation}
    \frac{d}{dr} \frac{(1-e^{-\Lambda(r)})}{2 r^2} = \frac{1}{\sqrt{1+4\alpha(\frac{2 m(r)}{r^3}-\frac{q(r)^2}{r^4})}} \frac{d}{dr} \left( \frac{m(r)}{r^3} - \frac{q(r)^2}{2 r^4}\right)
\end{equation}

The next step of the argument hinges on the assumption that the quantity $\frac{(1-e^{-\Lambda(r)})}{r^2}$ decreases as radial distance increases \cite{chakraborty_2020}. In the uncharged case this is equivalent to assuming decreasing mass density as the star's surface is approached (though we  note that some exotic star models do not have monotonically decreasing density profiles \cite{zhang_2021_stellar}).  
This assumption also seems reasonable in the case of a  charged sphere \cite{chakraborty_2020} and we employ it in what follows:

Utilizing the conservation equation
\begin{equation}
(\rho(r)+P(r))=-\frac{2}{\Phi'(r)}\left(\frac{\mathrm{d} P}{\mathrm{~d} r}-\frac{q}{4 \pi r^4} \frac{\mathrm{d} q}{\mathrm{~d} r}\right)
\end{equation}
we can write 
\begin{equation}\label{eq:dpdr}
\frac{dP}{dr} = \frac{q}{4 \pi r^4}\frac{dq}{dr}-\frac{1}{2}(\rho(r)+P(r))\Phi'(r).
\end{equation}
Recalling the $rr$ component of the field equations, we obtain
\begin{equation}\label{eq:rr}
\begin{aligned}
&\frac{1}{r^2}\left[r \Phi'(r) e^{-\Lambda(r)}-\left(1-e^{-\Lambda(r)}\right)\right] \\
&\qquad +\alpha \frac{\left(1-e^{-\Lambda(r)}\right)}{r^4}\left[2 r \Phi'(r) e^{-\Lambda(r)}+\left(1-e^{-\Lambda(r)}\right)\right]=\kappa \left(P(r)-\frac{q(r)^2 }{8 \pi r^4}\right)
\end{aligned}
\end{equation}
and subsequently get
\begin{equation}\label{eq:difflhs}
\begin{aligned}
 &\kappa \left(\frac{d P}{d r} + \frac{q(r)^2}{2 \pi r^5} - \frac{q(r)}{4 \pi r^4}\frac{d q}{d r}\right)=\kappa\left(\frac{q(r)^2}{2 \pi r^5}-\frac{1}{2}(\rho(r)+P(r))\Phi'(r)  \right)
\end{aligned}
\end{equation}
by differentiating the left hand side of  \eqref{eq:rr} using
\eqref{eq:dpdr}.
 
Addition of the $tt$ and $rr$ equations yields
\begin{equation}
(\rho(r)+P(r)) =\frac{1}{\kappa}\left(\frac{1}{r}e^{-\Lambda(r)}\left(\Lambda(r)^{\prime}+\Phi(r)^{\prime}\right)+\frac{2\alpha}{r^3} e^{-\Lambda(r)}\left(1-e^{-\Lambda(r)}\right)\left(\Lambda(r)^{\prime}+\Phi(r)^{\prime}\right)\right)
\end{equation}
after division by $r^2$. This  can be substituted into \eqref{eq:difflhs} to obtain
\begin{equation}
\begin{aligned}
 \kappa \left(\frac{d P}{d r} + \frac{q(r)^2}{2 \pi r^5} - \frac{q(r)}{4 \pi r^4}\frac{d q}{d r}\right)&=\kappa\frac{q(r)^2}{2 \pi r^5}-\frac{1}{2}\left(\frac{1}{r}e^{-\Lambda(r)} +\frac{2\alpha}{r^3} e^{-\Lambda(r)}\left(1-e^{-\Lambda(r)}\right) \right)\left(\Lambda(r)^{\prime}+\Phi(r)^{\prime}\right)\Phi'(r).
\end{aligned}
\end{equation}
Making use of equation \eqref{eq:rr} we find 
\begin{equation}
\begin{aligned}
&\frac{d}{dr}\left(\frac{1}{r^2}\left[r \Phi'(r) e^{-\Lambda(r)}-\left(1-e^{-\Lambda(r)}\right)\right]+\alpha \frac{\left(1-e^{-\Lambda(r)}\right)}{r^4}\left[2 r \Phi'(r) e^{-\Lambda(r)}+\left(1-e^{-\Lambda(r)}\right)\right]\right)\\
&=\kappa\frac{q(r)^2}{2 \pi r^5}-\frac{1}{2}\left(\frac{1}{r}e^{-\Lambda(r)} +\frac{2\alpha}{r^3} e^{-\Lambda(r)}\left(1-e^{-\Lambda(r)}\right) \right)\left(\Lambda(r)^{\prime}+\Phi(r)^{\prime}\right)\Phi'(r).
\end{aligned}
\end{equation}

If we define $\beta(r)=\frac{1-e^{-\Lambda(r)}}{r^2}$ the above can be rewritten as
\begin{equation}
\begin{aligned}
&\frac{d}{dr}\left(\frac{\Phi'(r)e^{-\Lambda(r)}}{r}\right)(1 + 2 \alpha \beta(r)) + Z(r) =\kappa\frac{q(r)^2}{2 \pi r^5}-\frac{\left(\Phi'(r)+\Lambda'(r)\right)}{2} \frac{\Phi'(r) e^{-\Lambda(r)}}{r}(1 + 2 \alpha \beta(r))
\end{aligned}
\end{equation}
where $Z(r) = \left(\frac{2 \alpha \Phi'(r) e^{-\Lambda(r)}}{r}-1+2\alpha \beta(r)\right) \beta'(r)$.
Equivalently,
\begin{align}
&2 e^{-(\Phi(r)+\Lambda(r))/2}\frac{d}{dr}\left[\frac{e^{-\Lambda(r)/2}}{r}\frac{d e^{\Phi(r)/2}}{dr}\right](1 + 2 \alpha \beta(r)) -\kappa\frac{q(r)^2}{2 \pi r^5}
\nonumber \\
& \qquad\qquad =\left(1-\frac{2 \alpha \Phi'(r) e^{-\Lambda(r)}}{r}-2\alpha \beta(r)\right) \beta'(r)
\label{eq43}
\end{align}
which can be shown using 
the identity
\begin{equation}
    \left(\frac{d}{dr}\left(\frac{\Phi'(r)e^{-\Lambda(r)}}{r} \right)+ \frac{1}{2} (\frac{\Phi'(r)e^{-\Lambda(r)}}{r})(\Phi'(r)+\Lambda'(r))\right)=2 e^{-(\Phi(r)+\Lambda(r))/2}\frac{d}{dr}\left[\frac{e^{-\Lambda(r)/2}}{r}\frac{d e^{\Phi(r)/2}}{dr}\right].
\end{equation}

To proceed further we assume the factors $(1+2\alpha \beta(r))$ and $\left(1-\frac{2 \alpha \Phi'(r) e^{-\Lambda(r)}}{r}-2\alpha \beta(r)\right)$ are both non-negative, which is valid for
sufficiently small, positive $\alpha$. The right hand side of \eqref{eq43} is then  negative due to the monotonically decreasing nature of $\beta(r)$. We can thus write  
\begin{equation} \label{eq:31}
    2 e^{-(\Phi(r)+\Lambda(r))/2}\frac{d}{dr}\left[\frac{e^{-\Lambda(r)/2}}{r}\frac{d e^{\Phi(r)/2}}{dr}\right](1 + 2 \alpha \beta(r)) -\kappa\frac{q(r)^2}{2 \pi r^5}\leq 0
\end{equation}
which upon integration yields
\begin{equation}
e^{\Phi(r'')/2} \leq \kappa \int^{r''}  dr' r'e^{\Lambda(r')/2}\int^{r'} dr
e^{(\Phi(r)+\Lambda(r))/2} \frac{q(r)^2}{4 \pi r^5}(1 + 2 \alpha \beta(r))^{-1} \; .
\end{equation}
Making for convenience    the following definitions 
\begin{equation}
\begin{aligned}
\mathcal{Q} &\equiv e^{(\Phi(r)+\Lambda(r))/2} \frac{q(r)^2}{4 \pi r^5}(1 + 2 \alpha \beta(r))^{-1}\\
   \eta(r'') &\equiv  \int^{r''} e^{\Lambda(r')/2}r'  \int^{r'}\mathcal{Q}(r) dr dr' \\
   \psi &\equiv e^{\Phi(r'')/2}- \eta(r'') \qquad
   \xi \equiv \int^{r''} r^{\prime} e^{\Lambda(r')/2} d r^{\prime}  
\end{aligned}
\end{equation}
 it is straightforward to show that
\begin{align}
&\frac{d \psi}{d \xi} = \frac{1}{r''}e^{-\Lambda(r'')/2}\frac{d}{dr''} e^{\Phi(r'')/2} - \int^{r''} dr \mathcal{Q}(r'),
\\
&\frac{d^2 \psi}{d \xi^2} = \frac{1}{r''}e^{-\Lambda(r'')/2} \frac{d}{dr}\left[\frac{1}{r''} e^{-\Lambda(r'')/2} \frac{d}{dr''}e^{\Phi(r'')/2}\right] - \frac{1}{r''}e^{-\Lambda(r'')/2}\mathcal{Q}(r'') \leq 0
\end{align}
where integration is from the center ($r''>r'>r$), 
or alternatively
\begin{equation}
     \frac{d^2 \psi}{d \xi^2} r e^{\Lambda(r)/2} \leq 0
\end{equation}
which follows from  \eqref{eq:31}.  
Since $e^{\Lambda(r)/2}>0$ and $r>0$ this is equivalent to 
\begin{equation}
     \frac{d^2 \psi}{d \xi^2} \leq 0.
\end{equation}
Trivially integrating and applying the mean value theorem \cite{bohmer_2007} we see that
\begin{equation}
\frac{d \psi}{d \xi} \leq \frac{\psi(\xi)-\psi(0)}{\xi-0} \leq \frac{\psi(\xi)}{\xi}
\end{equation}
since at the center of the star
 $\xi=0$ and $\psi(0)>0$ since 
 $e^{\Phi(r)/2}>0$ inside the star.
 
Putting everything together  we find
\begin{align}\label{eq:plugged}
&\left(\frac{1}{r}e^{-\Lambda(r'')/2}\frac{d}{dr} e^{\Phi(r'')/2} - \int^{r''} dr' \mathcal{Q}(r')\right) \left( \int^{r''} d r^{\prime} r^{\prime} e^{\Lambda(r') / 2} \right) \nonumber\\
&\qquad\qquad\qquad \leq e^{\Phi(r'')/2}-\int^{r''} r' e^{\Lambda_{\mathrm{int}} / 2}  \int^{r'}\mathcal{Q}(r) dr dr'.
\end{align}

Recalling that $\beta(r) = \frac{1-e^{-\Lambda(r)}}{r^2}=-\frac{1}{2 \alpha}\left[1-\sqrt{1+\frac{\kappa \alpha}{2 \pi}\left(\frac{2 m(r)}{r^3}-\frac{q(r)^2}{r^4}\right)}\right]$ such that
\begin{equation}
    e^{-\Lambda(r)}=1-r^2\beta(r),
\end{equation}
and that $\beta(r)\geq\beta(r')\geq\beta(r'')$ (more primes corresponding to further from the center), we obtain
\begin{equation}
\begin{aligned}
\xi \equiv \int^{r''} d r^{\prime} r^{\prime} e^{\Lambda(r') / 2} = \int^{r''} d r^{\prime} r^{\prime} \frac{1}{\sqrt{1-r'^2\beta(r')}} & \geq \int^{r''} d r^{\prime} r^{\prime} \frac{1}{\sqrt{1-r'^2\beta(r'')}} \\
& = \frac{1}{\beta(r'')}\left(1 - \sqrt{1-r''^2 \beta(r'')}\right)
\end{aligned}
\end{equation}
as a bound  on $\xi$. 

Similarly we can consider the term
\begin{equation}
\begin{aligned}
    \int^{r''}\mathcal{Q}(r') dr &=\kappa \int^{r''} e^{(\Phi(r')+\Lambda(r'))/2} \frac{q(r')^2}{4 \pi r'^5}(1 + 2 \alpha \beta(r'))^{-1} dr\\
    &=\frac{\kappa}{4 \pi} \int^{r''} \frac{e^{\Phi(r') / 2}}{\sqrt{1 - r'^2 \beta(r')}\left(1+2\alpha\beta(r') \right)} \frac{q(r')^2}{ r'^5}  dr'.
\end{aligned}
\end{equation}
Combining our previous assumption on $\beta$ with the following assumption regarding the behaviour of the charge:
\begin{equation}
    e^{\Phi_{\mathrm{int}}(r) / 2}\frac{q(r)^2}{ r^5} \geq e^{\Phi_{\mathrm{int}}(r') / 2}\frac{q(r')^2}{ r'^5} \geq e^{\Phi_{\mathrm{int}}(r'') / 2}\frac{q(r'')^2}{ r''^5}
\end{equation}
the above integral can be bounded as follows:
\begin{equation}
\begin{aligned}
    \frac{\kappa}{4 \pi} \int^{r''} \frac{e^{\Phi(r') / 2}}{\sqrt{1 - r'^2 \beta(r')}\left(1+2\alpha\beta(r') \right)} \frac{q(r')^2}{ r'^5}  dr' &\geq \frac{\kappa}{4 \pi} \frac{e^{\Phi_{\mathrm{int}}(r'') / 2}}{\left(1+2\alpha\beta(0) \right)}\frac{q(r'')^2}{ r''^5} \int^{r''} \frac{dr'}{\sqrt{1 - r'^2 \beta(r'')}} \\
    &=\frac{\kappa}{4 \pi} \frac{e^{\Phi_{\mathrm{int}}(r'') / 2}}{\left(1+2\alpha\beta(0) \right)}\frac{q(r'')^2}{ r''^5} \frac{\mathrm{arcsin}\left(\sqrt{\beta(r'')} r''\right)}{\sqrt{\beta(r'')}}.
\end{aligned}
\end{equation}

Finally we can consider the most complicated integral term:
\begin{equation}
\begin{aligned}
    &\int^{r''} r' e^{\Lambda_{\mathrm{int}} / 2}  \int^{r'}\mathcal{Q}(r) dr dr' = \frac{\kappa}{4 \pi}\int^{r''} r' e^{\Lambda_{\mathrm{int}} / 2} \int^{r'} \frac{e^{\Phi_{\mathrm{int}} / 2}}{\sqrt{1 - r^2 \beta(r)}\left(1+2\alpha\beta(r) \right)} \frac{q(r)^2}{ r^5}  dr dr'\\
    &\geq \frac{\kappa}{4 \pi} \left(1+2\alpha\beta(0) \right)^{-1}\int^{r''} r'  e^{(\Phi_{\mathrm{int}}(r')+\Lambda_{\mathrm{int}}(r')) / 2}\frac{q(r')^2}{ r'^5} \frac{\mathrm{arcsin}\left(\sqrt{\beta(r')} r'\right)}{\sqrt{\beta(r')}} dr'\\
    &\geq \frac{\kappa}{4 \pi} \left(1+2\alpha\beta(0) \right)^{-1}e^{\Phi_{\mathrm{int}}(r'') / 2}\frac{q(r'')^2}{ r''^5} \int^{r''} r'^2  \frac{1}{\sqrt{1-r'^2 \beta(r')}} \frac{\mathrm{arcsin}\left(\sqrt{\beta(r')} r'\right)}{\sqrt{\beta(r')}r'} dr'.
\end{aligned}
\end{equation}
Since  the function $\mathrm{arcsin}(x)/x$ monotonically increases on the interval $x \in [0,1]$, the integral is minimal when the smallest $\beta$ is chosen. Thus
\begin{equation}
\begin{aligned}
    &\int^{r''} r' e^{\Lambda_{\mathrm{int}} / 2}  \int^{r'}\mathcal{Q}(r) dr dr' \\
    &\geq \frac{\kappa}{4 \pi} \left(1+2\alpha\beta(0) \right)^{-1}e^{\Phi_{\mathrm{int}}(r'') / 2}\frac{q(r'')^2}{ r''^5} \int^{r''} r'^2  \frac{1}{\sqrt{1-r'^2 \beta(r'')}} \frac{\mathrm{arcsin}\left(\sqrt{\beta(r'')} r'\right)}{\sqrt{\beta(r'')}r'} dr'\\
    &=\frac{\kappa}{4 \pi} \left(1+2\alpha\beta(0) \right)^{-1}e^{\Phi_{\mathrm{int}}(r'') / 2}\frac{q(r'')^2}{ r''^5} \frac{1}{\beta(r'')}\left(r'' - \sqrt{\frac{1 - \beta(r'')r''^2}{\beta(r'')}}\mathrm{arcsin}(\sqrt{\beta(r'')}r'') \right).
\end{aligned}
\end{equation}
Inserting these results into \eqref{eq:plugged}
yields
\begin{equation}
\begin{aligned}
    &\left(e^{(\Phi(r'')-\Lambda(r''))/2}\frac{\Phi'(r'')}{2r} - \frac{\kappa}{4 \pi} \frac{e^{\Phi(r'') / 2}}{\left(1+2\alpha\beta(0) \right)}\frac{q(r'')^2}{ r''^5} \frac{\mathrm{arcsin}\left(\sqrt{\beta(r'')} r''\right)}{\sqrt{\beta(r'')}}\right) \left(\frac{\left(1 - \sqrt{1-r''^2 \beta(r'')}\right)}{\beta(r'')} \right) 
   \\
    &\qquad \leq e^{\Phi(r'')/2}-\frac{\kappa}{4 \pi} \left(1+2\alpha\beta(0) \right)^{-1}e^{\Phi(r'') / 2}\frac{q(r'')^2}{ r''^5} \frac{1}{\beta(r'')}\left(r'' - \sqrt{\frac{1 - \beta(r'')r''^2}{\beta(r'')}}\mathrm{arcsin}(\sqrt{\beta(r'')}r'') \right)
\end{aligned}
\end{equation}
and  cancelling a factor of $e^{\Phi(r'')/2}$ leaves
\begin{equation}
\begin{aligned}
    &\left(\frac{\sqrt{1-r''^2\beta(r'')}}{r''}\frac{\Phi'(r'')}{2} - \frac{\kappa}{4 \pi} \frac{1}{\left(1+2\alpha\beta(0) \right)}\frac{q(r'')^2}{ r''^5} \frac{\mathrm{arcsin}\left(\sqrt{\beta(r'')} r''\right)}{\sqrt{\beta(r'')}}\right) \left(\frac{\left(1 - \sqrt{1-r''^2 \beta(r'')}\right)}{\beta(r'')} \right) \\
    & \qquad\qquad \leq
    1-\frac{\kappa}{4 \pi} \left(1+2\alpha\beta(0) \right)^{-1}\frac{q(r'')^2}{ r''^5} \frac{1}{\beta(r'')}\left(r'' - \sqrt{\frac{1 - \beta(r'')r''^2}{\beta(r'')}}\mathrm{arcsin}(\sqrt{\beta(r'')}r'') \right)
\end{aligned}
\end{equation}
which depends explicitly on $\Phi'(r'')$. This can be determined from the $rr$ field equation \eqref{eq:ferr} to give

\begin{equation}
\Phi'(r) = \frac{\alpha +e^{2 \Lambda(r)} \left(\alpha -8 \pi  r^4 P(r)+q(r)^2-r^2\right)+e^{\Lambda(r)} \left(r^2-2 \alpha \right)}{2 \alpha  r-r e^{\Lambda(r)} \left(2 \alpha +r^2\right)}.
\end{equation}
Making this substitution and simplifying, we find the desired inequality 
\begin{equation}
\begin{aligned}
    &\frac{\left(\sqrt{1-R^2 \beta (R)}-1\right) \left(q(R)^2+R^4 \beta (R)(-1+\alpha \beta(R))\right)}{2 R^4 \beta (R) \sqrt{1-R^2 \beta (R)}(1+2 \alpha \beta(R))} \\
    &\qquad\qquad \qquad \leq  1+\frac{2 q(r)^2}{R^4 \beta(R)}\left(\frac{\mathrm{arcsin}( \sqrt{\beta(R)}R)}{\sqrt{\beta(R)}R}-1\right)
\end{aligned}
\label{genBuch}
\end{equation}
which is the generalized Buchdahl bound for charged spheres in 4DEGB gravity. 

In the $q\to 0$ limit we recover the following inequality \cite{chakraborty_2020}
\begin{equation}
\sqrt{1-\beta(R) R^2}\left(1+ \alpha \beta(R)\right)>\frac{1}{3}\left(1- \alpha \beta(R)\right) 
\end{equation}
for uncharged objects in 4DEGB.

Alternatively, in the small $\alpha$ limit  $\beta_\mathrm{\alpha \to 0}(r) = \frac{2 m(r)}{r^3} - \frac{q(r)^2}{r^4}$, and consequently
\begin{equation}
    \frac{\left(\sqrt{1-r^2 \beta (r)}-1\right) \left(q(r)^2-r m(r)\right)}{r^4 \sqrt{1-r^2 \beta (r)}}\leq\beta (r)+\frac{2 q(r)^2 }{r^4}\left(\frac{\sin ^{-1}\left(\sqrt{r^2 \beta (r)}\right)}{\sqrt{r^2 \beta (r)}}-1\right)
\end{equation}
which matches\footnote{At the time of publication \cite{bohmer_2007} has a factor of 2 missing from their final expression which Dr. Harko has kindly confirmed to be a typo.
We note that in  \cite{bohmer_2007} $\alpha(r)=\beta_\mathrm{\alpha \to 0}(r)\frac{r^3}{2 m(r)}$ in our notation.
} the final inequality from \cite{bohmer_2007}. 

\subsection{Relation to the Black Hole Horizon}

For a given $\alpha$ 
the radius of a black hole is
$R=M+\sqrt{M^2-Q^2-\alpha}$. Since the minimum mass  $M_\mathrm{min}=\sqrt{Q^2+\alpha}$, we obtain $R_\mathrm{Mmin}=\sqrt{Q^2+\alpha}$.
Substituting  this into the Buchdahl bound 
\eqref{genBuch}
we find that it is automatically satisfied as an equality; in other words the Buchdahl bound intersects the minimum mass point of the black hole horizon. The $M/R$ curves for nonzero $\alpha$ also smoothly join this point in the limit of large central pressure, as can be seen in figures \ref{fig:results1} and \ref{fig:results2}.

\section{Results }\label{sec:results}
\begin{figure*}
        \subfloat[\label{fig:GR mr}]{
        \includegraphics[width=7.6cm]{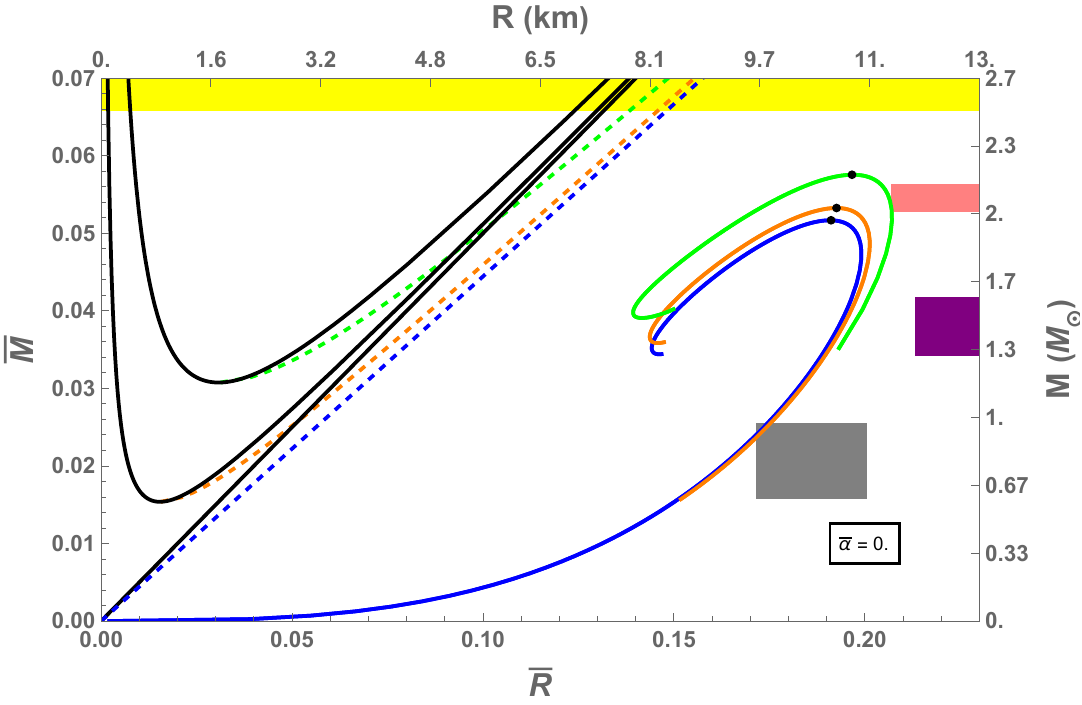}
        }\hfill
        \subfloat[\label{fig:GR mpc}]{
        \includegraphics[width=7.6cm]{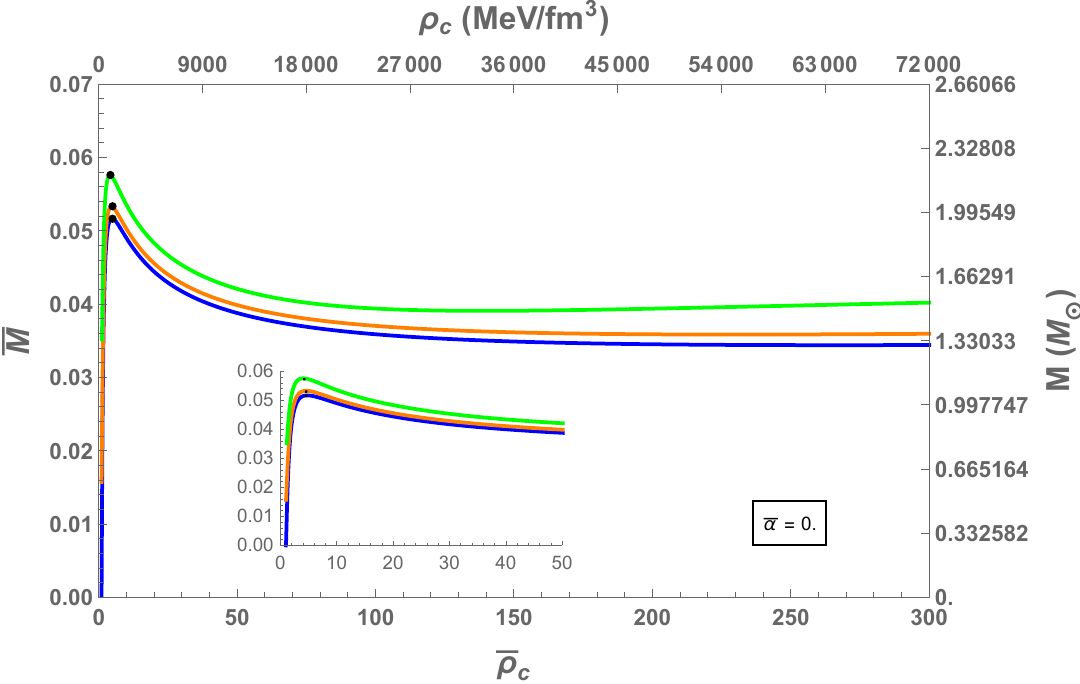}
        }

        \subfloat[\label{fig:apt0005 mr}]{
        \includegraphics[width=7.6cm]{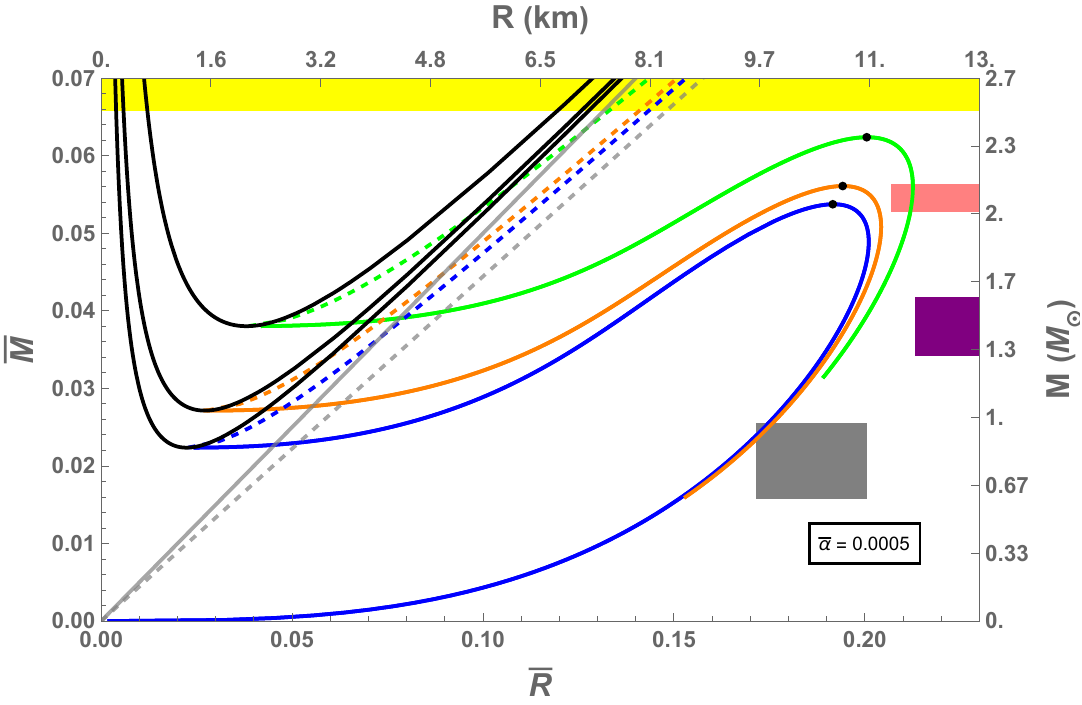}
        }\hfill
        \subfloat[\label{fig:apt0005 mpc}]{
        \includegraphics[width=7.6cm]{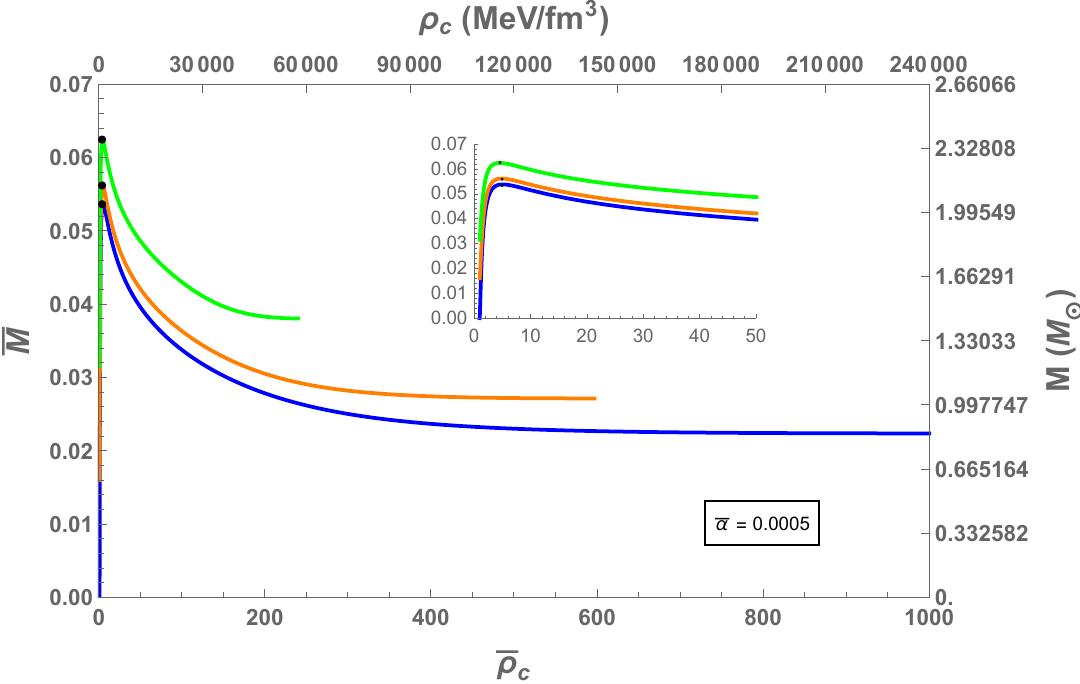}
        }

        \subfloat[\label{fig:apt001 mr}]{
        \includegraphics[width=7.6cm]{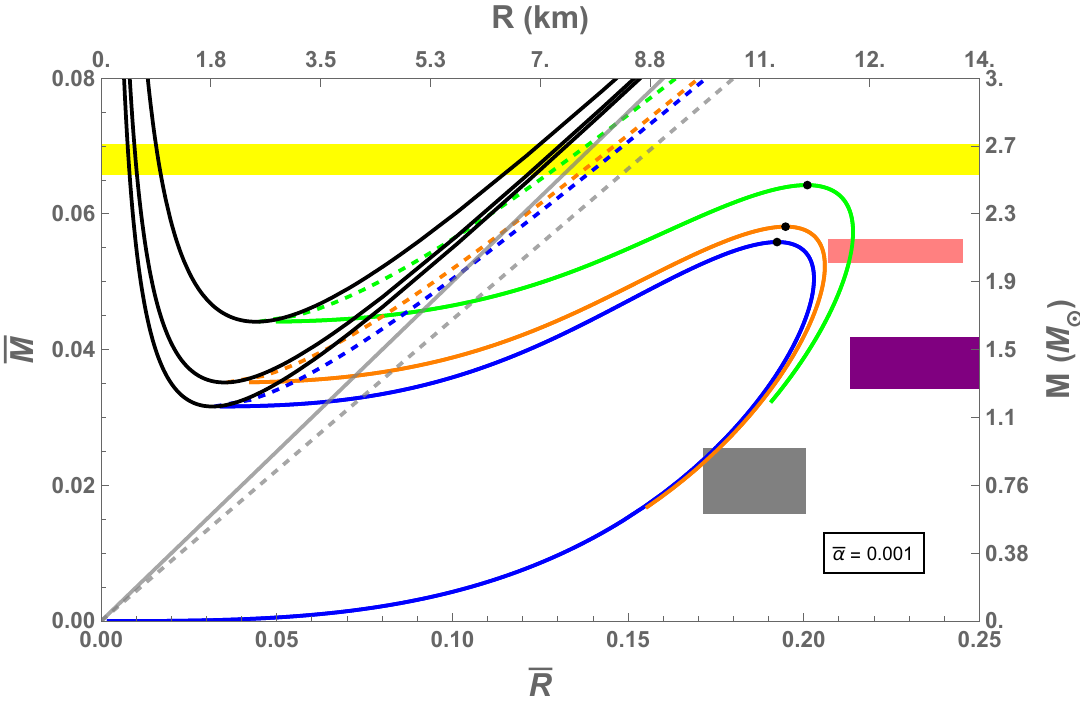}
        }\hfill
        \subfloat[\label{fig:apt001 mpc}]{
        \includegraphics[width=7.6cm]{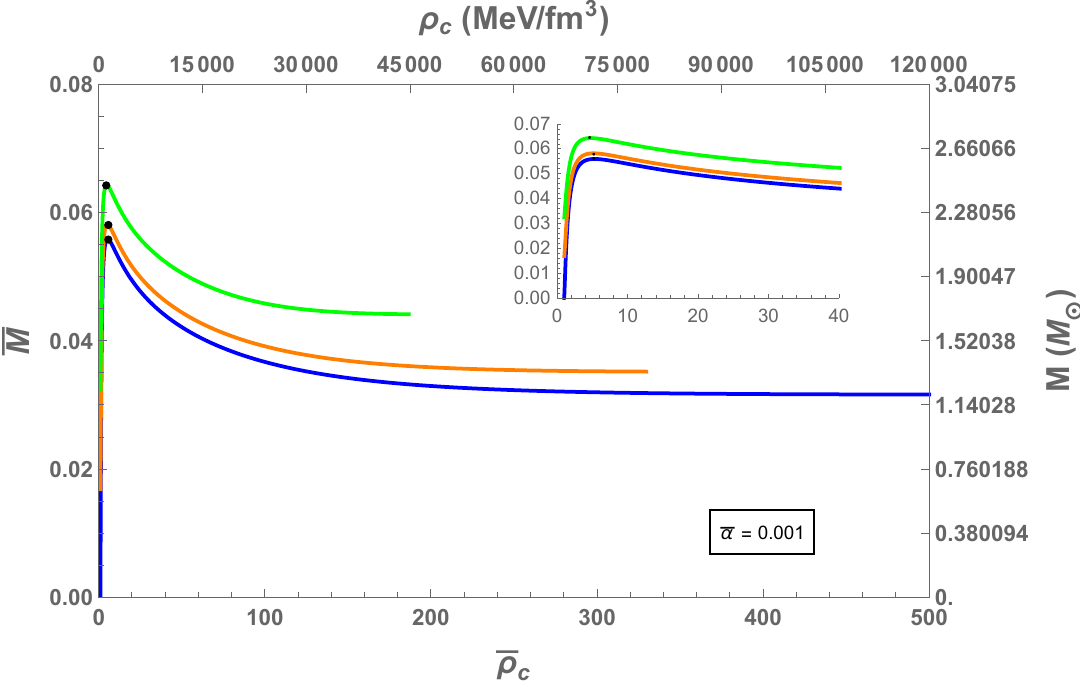}
        }

	\caption[]{Mass vs. radius/mass vs. central density curves for charged 4DEGB quark stars for three different fixed charges. The (blue, orange, green) curves correspond to charges $\bar{Q}=(0, 1.538, 3.076)\times10^{-2}$ respectively, with black dots corresponding to local maximum mass points (when present). Note that the orange and green curves begin at a nonzero value of $\overline{R}$. The grey solid and dashed lines are the uncharged GR Schwarzschild and Buchdahl bounds, respectively. The coloured dashed lines are the 4DEGB Buchdahl bounds for the three different charges, with the colours corresponding to the associated $M/R$ curve. The black curves are the 4DEGB black hole horizons, with charges matching the Buchdahl/$MR$ curves that intersect them. Finally, the coloured boxes are $1\sigma$ observational estimates of mass/radii for PSR J0030+0451 (purple) \cite{Miller_2019}, PSR J0740+6620 (pink) \cite{salmi2024}, HESS J1731-347 (gray) \cite{Horvath_2023}, and GW190814 (yellow) \cite{Abbott_2020}.   \label{fig:results1}}
\end{figure*}

\begin{figure*}

        \subfloat[\label{fig:apt01 mr}]{
        \includegraphics[width=7.6cm]{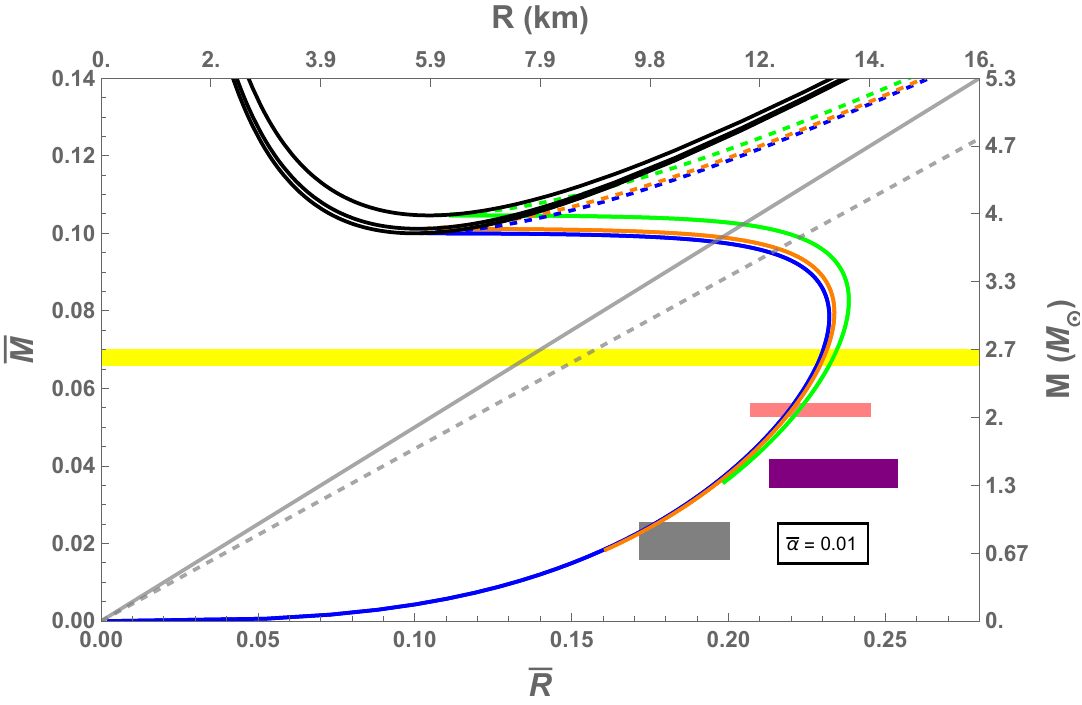}
        }\hfill
        \subfloat[\label{fig:apt01 mpc}]{
        \includegraphics[width=7.6cm]{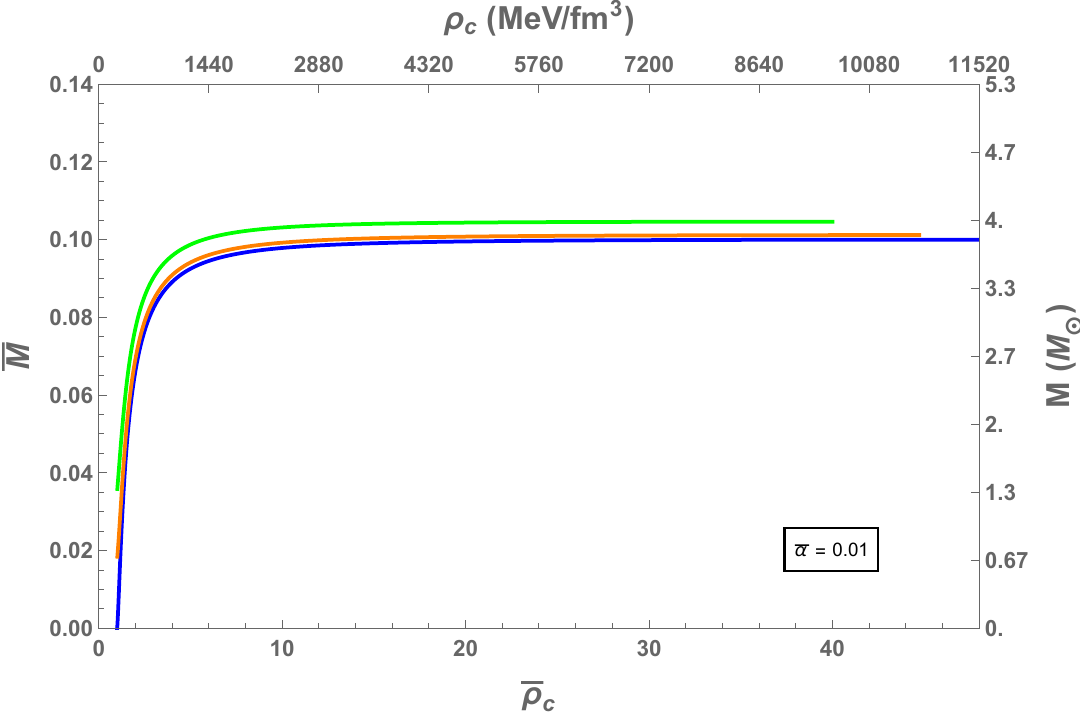}
        }

        \subfloat[\label{fig:apt05 mr}]{
        \includegraphics[width=7.6cm]{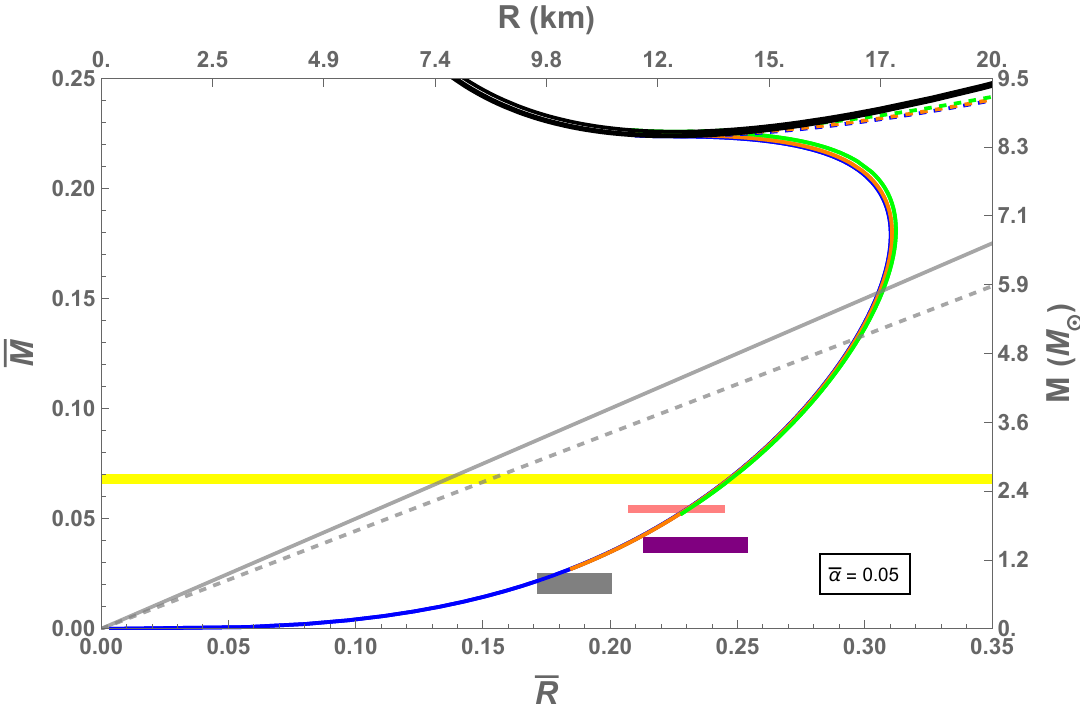}
        }\hfill
        \subfloat[\label{fig:apt05 mpc}]{
        \includegraphics[width=7.6cm]{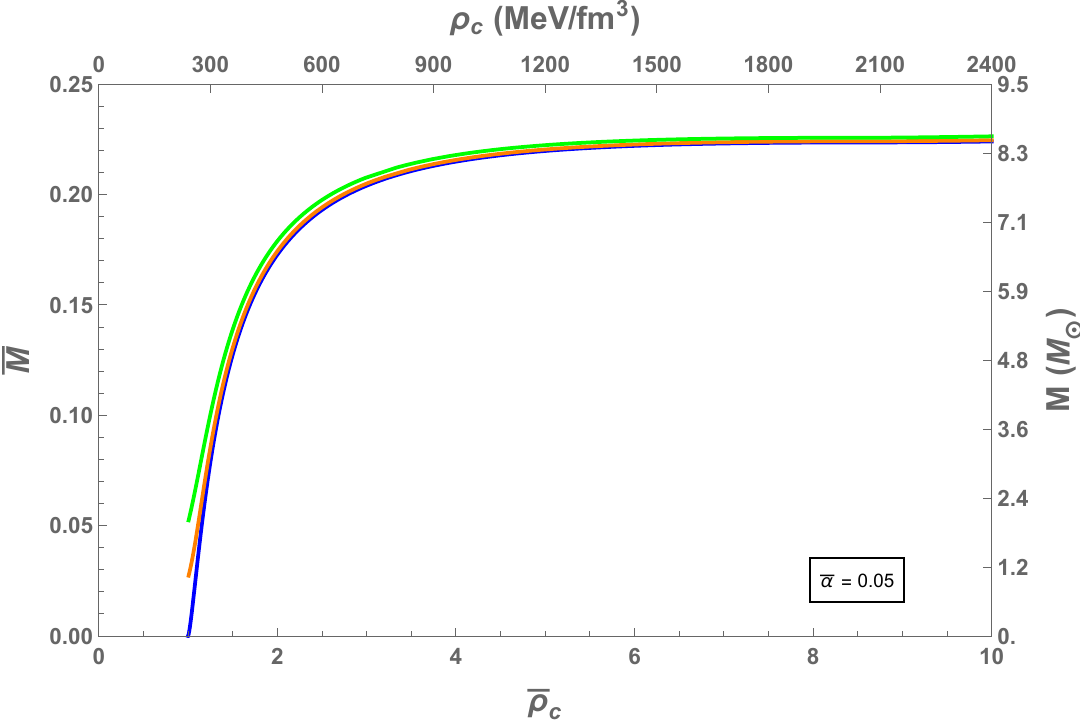}
        }

        \subfloat[\label{fig:apt1 mr}]{
        \includegraphics[width=7.6cm]{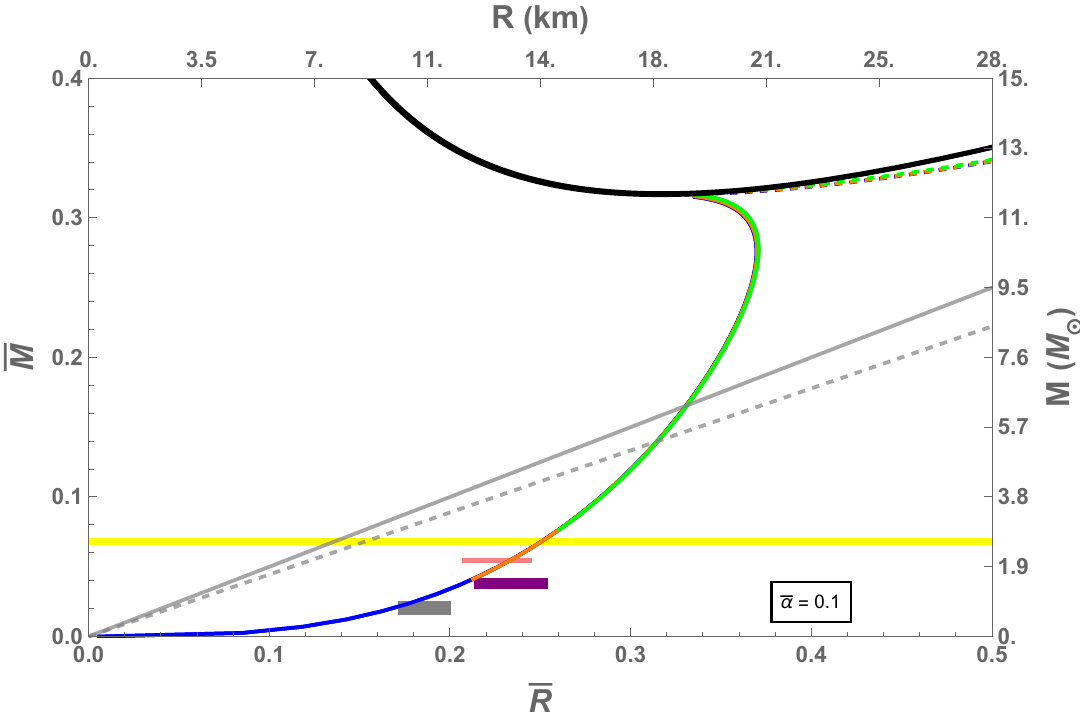}
        }\hfill
        \subfloat[\label{fig:apt1 mpc}]{
        \includegraphics[width=7.6cm]{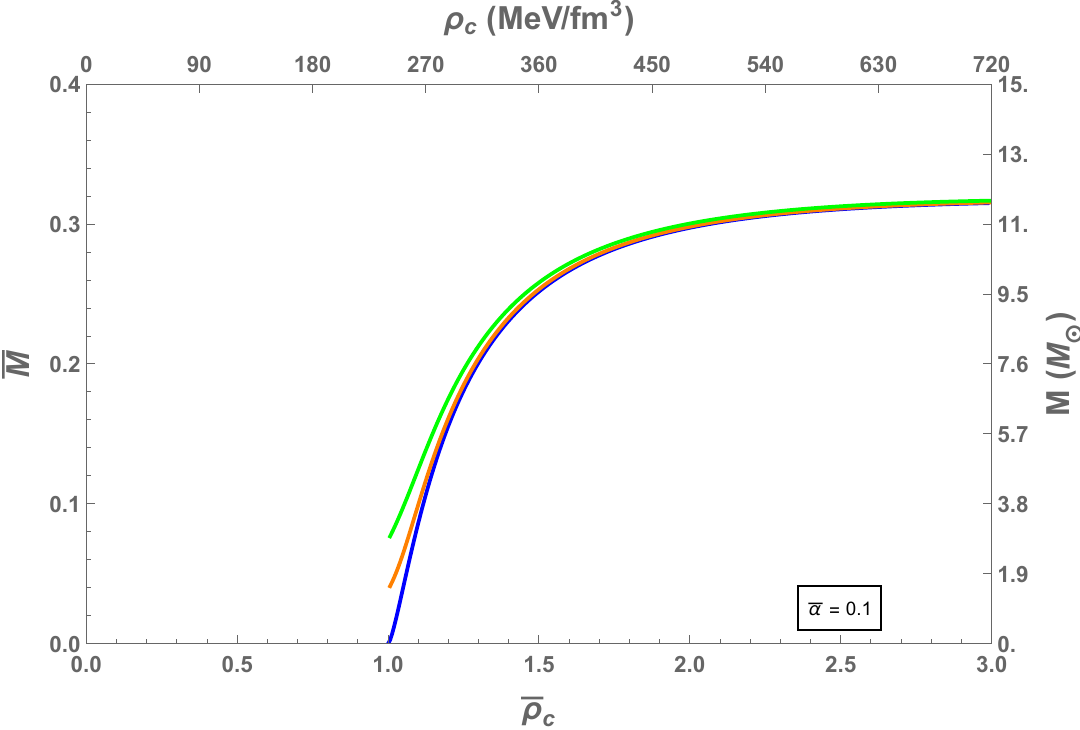}
        }

	\caption[]{Mass vs. radius/mass vs. central density curves for charged 4DEGB quark stars for three different fixed charges. The (blue, orange, green) curves correspond to charges $\bar{Q}=(0, 1.538, 3.076)\times10^{-2}$ respectively, with black dots corresponding to local maximum mass points (when present). Note that the orange and green curves begin at a nonzero value of $\overline{R}$ and almost perfectly overlap with the blue curve at larger $\alpha$.
 The grey solid and dashed lines are the uncharged GR Schwarzschild and Buchdahl bounds, respectively. The coloured dashed lines are the 4DEGB Buchdahl bounds for the three different charges, with the colours corresponding to the associated $M/R$ curve. The black curves are the 4DEGB black hole horizons, with charges matching the Buchdahl/$MR$ curves which intersect them. Finally, the coloured boxes are $1\sigma$ observational estimates of mass/radii for PSR J0030+0451 (purple) \cite{Miller_2019}, PSR J0740+6620 (pink) \cite{salmi2024}, HESS J1731-347 (gray) \cite{Horvath_2023}, and GW190814 (yellow) \cite{Abbott_2020}.   \label{fig:results2}}
\end{figure*}
The mass/radius ($M/R$) and mass/central density ($M/\rho_c$) curves for charged 4DEGB quark stars are presented in figures \ref{fig:results1} and \ref{fig:results2}. As expected from previous work
\cite{gammon_2024,zhang_2021_stellar}, 
we observe the general trends that for  a larger $\alpha$ and/or $Q$,   the mass/radius profiles of the quark stars increase in size. For a fixed nonzero charge the stars have a minimum size, 
below which the gravitational attraction cannot overcome the self-repulsion of the charge. This is most evident in the right-hand lower panels of
figure~\ref{fig:results2}. 
As $\alpha$ gets large the $M/R$ curves converge for all values of fixed $Q$ (and similarly with the $M/\rho_c$ curves)  with a slight divergence near $\bar{\rho_c}=1$ in the $M/\rho_c$ plots to account for the structures mentioned above.

As in \cite{gammon_2024},   even for small, nonzero $\alpha$ we find curves for the Buchdahl bound that intersect the black hole horizon at the minimum mass point. A new feature, evidently not previously noted,  is that this same interaction between the black hole horizon and Buchdahl bound can be observed in GR when $Q \neq 0$. This is evident in the orange and green dashed curves in the upper left panel of figure~\ref{fig:results1}. 
 The difference is that for nonzero $\alpha$ the $M/R$  curves approach this intersection point smoothly, whereas in GR the curves turn away from the horizon/BB and thus never meet. 

On the solution curves we have indicated maximum mass points (when this point doesn't occur at the intersection with the horizon) with a black dot - in general relativity these maximum mass points indicate a transition point from stability against radial perturbations to instability for uncharged stars. In 4DEGB it's not clear whether this coincidence holds, and offers an interesting avenue for future research.

\begin{figure}
    \centering
    \includegraphics[width=0.5\linewidth]{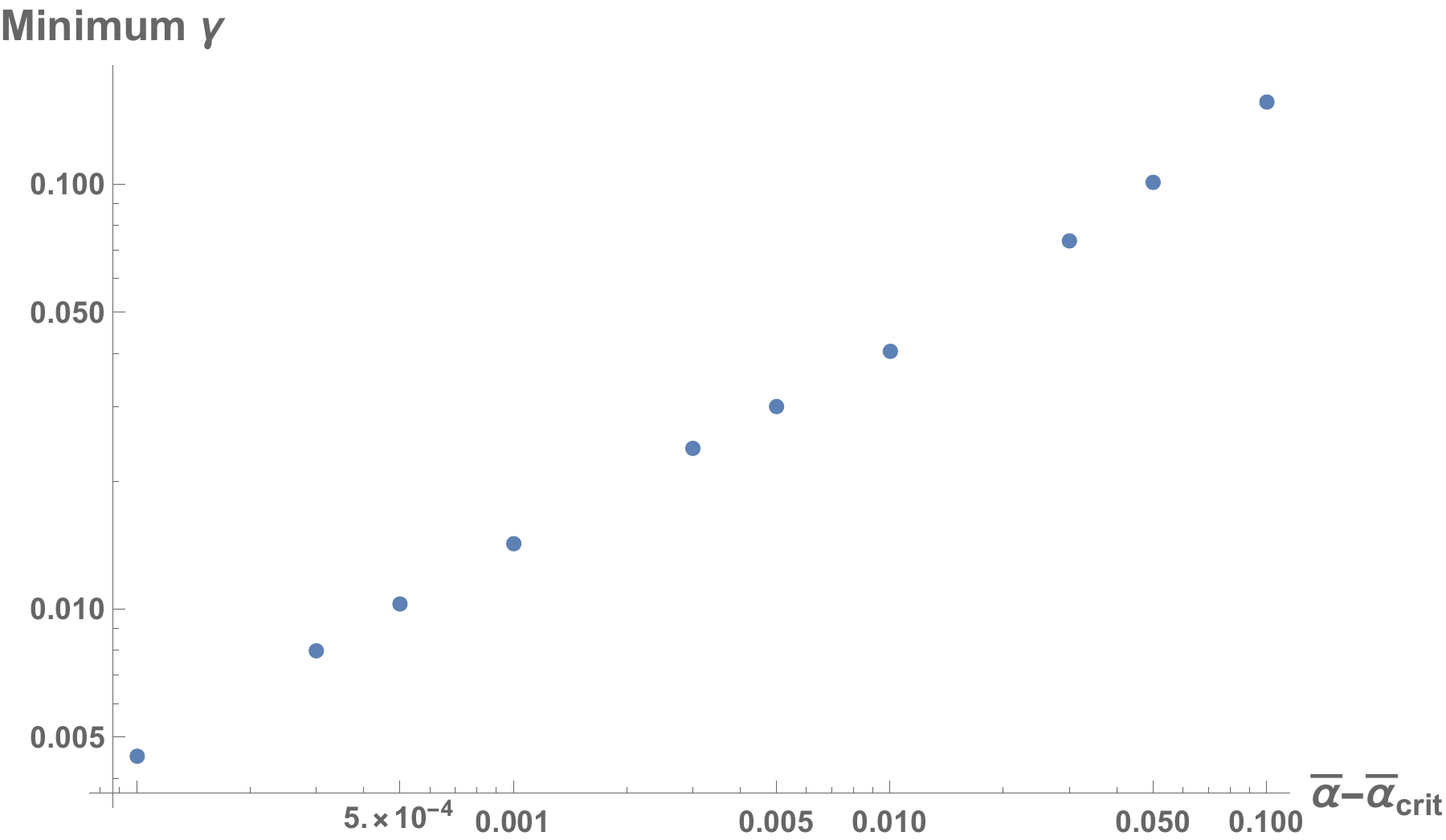}
    \caption{In the above figure we numerically plot
    the minimal value of
    $\gamma$ for  $\bar{\alpha} > \bar{\alpha}_\mathrm{crit}$ and a fixed central pressure of $\bar{P}_0 = 0.00001$. 
    For a given 
    $\bar{\alpha} - \bar{\alpha}_\mathrm{crit}$ the
    charge parameter $\gamma$ must be larger than the 
    value
    indicated above 
    in order that the pressure  not diverge (a requirement for a physical stellar object with a finite radius).}
    \label{fig:minimumgamma}
\end{figure}

In the uncharged case a criticality condition was previously noted \cite{gammon_2024} wherein for $\bar{\alpha}\geq\frac{3}{4 \pi}$ a critical central pressure was present - for central pressures below this critical value, the pressure function diverges with no real roots to define a star's surface. Above this critical pressure we find numerical solutions which lie at or beyond the black hole horizon, indicating a lack of stable physical stars in this critical regime. Deriving an analogous bound for the charged case is considerably more complicated. However numerically we find that  for $Q > 0$ (and hence $\gamma>0$), such criticality can be pushed to $\bar{\alpha} > \frac{3}{4\pi}$. Figure \ref{fig:minimumgamma} shows a rough numerical sketch of the parameter space, illustrating the required value of the charge parameter $\gamma$ for solutions to describe physical stars when $\bar{\alpha} > \bar{\alpha}_\mathrm{crit}$.   Because of this lack of an analytic bound for the charged case, we do not present numerical solutions in which such behaviour is manifest.  With this, we note that in this paper we have considered values of $\bar{\alpha} \leq 0.1$. However, the basic behaviour of these edge cases is shown for the uncharged star in \cite{gammon_2024}, and at $\bar{\alpha} \gtrapprox 0.05$ the differences between the charges considered in this paper are heavily suppressed. Making $\bar{\alpha}$ appreciably larger than what we have considered  will lead to the emergence of  `critical central pressure' behaviour, with solutions lying on/past the black hole horizon  - examination of these cases requires a more detailed stability analysis, which we leave for future investigation.

Most interestingly we note, similar to the uncharged case
\cite{gammon_2024}, that the stars described by this theory can be ECCOs (Extreme Compact Charged Objects): charged stellar objects that are smaller (more compact) than the uncharged Buchdahl bound (and sometimes even the Schwarzschild radius) of general relativity.  For large enough $\alpha$ the associated $M/R$ curves do not reach their maximum mass until they intersect the black hole horizon - these are particularly interesting candidates for stable ECCOs since in pure uncharged GR we only see stability against radial perturbations when $dM/d\rho >0$. However, a net charge offsets the stability point from this maximum mass point \cite{zhang_2021_stellar} - it is likely that the modifications to gravity will have a similar offsetting effect which should be investigated in future studies before declaring these solutions stable against perturbations.

We close this section by briefly commenting on a comparison of our results to real candidate quark stars.  In the left column of figures~\ref{fig:results1} and \ref{fig:results2} we have included 
the $1\sigma$ mass/radius bounds for
HESS J1731-347   \cite{Horvath_2023},
PSR J0030+0451 \cite{Miller_2019}, PSR J0740+6620   \cite{salmi2024},  and GW190814  \cite{Abbott_2020}.  The
HESS J1731-347 object is particularly interesting due to its low mass, since its
formation is not thought to be viable within standard Stellar Evolution \cite{Suwa_2018}. We see that for sufficiently large $\bar{\alpha} \approx 0.05$, the simple non-interacting quark star equation of state \eqref{eq:eos} is consistent with known observations of these objects having various charges, whereas this is not true for the GR case. Since it is known from past work \cite{zhang_2021_unified,zhang_2021_stellar} that the inclusion of strong interaction effects can significantly inflate the mass-radius profiles of these stars, the smaller $\alpha$ solution sets could likely also fit this data with an interacting EOS. A more complete quantitative analysis of the observational constraints on the ($Q$, $\alpha$) parameter space that includes this more realistic equation of state for quark matter \cite{zhang_2021_stellar,zhang_2021_unified} is a worthwhile subject for future study.

\section{Stability}\label{sec:stability}

We now briefly consider the stability of stars respecting the generalized  Buchdahl bound. In general relativity a necessary but insufficient condition for an uncharged compact star to be stable against radial perturbations is $dM/d\rho_c>0$ \cite{zhang_2021_stellar,glendenningbook,arbanil2015}, corresponding to the part of the solution curve before a maximal mass point is reached. In Einstein-Maxwell theory a net charge has been shown to offset the stability point from the maximum mass point in either direction \cite{zhang_2021_stellar}. In a similar vein, when the coupling to   higher curvature gravity is nonzero, it is not obvious whether this coincidence of stability and the maximum mass point will hold for uncharged stars. While we leave a thorough analysis of the fundamental radial oscillation modes for future work (where stability against radial oscillations can be ensured), the speed of sound and effective adiabatic index inside the star can still be discussed.

In the interior of a stable star, the sound speed $c_s$ must never exceed the speed of light $c$. The non-interacting quark equation of state   \eqref{eq:eos} has a constant subluminal sound speed of $c_s = \frac{c}{\sqrt{3}}$, and thus the causality condition is always satisfied. Similarly, the effective adiabatic index 
\begin{equation}
\gamma_{\mathrm{eff}} \equiv\left(1+\frac{\rho}{P}\right)\left(\frac{d P}{d \rho}\right)_S
\end{equation}
is often used as another indicator of stability as it is seen as a bridge between ``the relativistic structure of a spherical static object and the equation of state of the interior fluid" \cite{moustakidis2017_stability}. The subscript $S$ in the above equation indicates that we consider the sound speed at a constant specific entropy. In principle a critical value for $\langle \gamma_\mathrm{eff} \rangle$ exists, below which configurations are unstable against radial perturbations. In standard general relativity this critical value can be written \cite{chandrasekhar1964,moustakidis2017_stability} $\gamma_{c r}=\frac{4}{3}+\frac{19}{42} \beta$, where $\beta=2 M / R = R_S/R$ is the compactness parameter. If $\beta \to 0$ the well-known classical Newtonian limit is recovered as expected ($\langle\gamma_\mathrm{eff}\rangle \geq \frac{4}{3}$).

An equivalent bound has not yet been derived for the 4DEGB theory. Despite this, it is common practice to plot the adiabatic index of the star relative to the Newtonian critical value \cite{banerjee_2021_quark,banerjee_2021_strange,hansraj_2020_isotropic,singh_2022_anisotropic}. Since we restrict ourselves to a simple, non-interacting equation of state \eqref{eq:eos} it is straightforward to show that $\gamma_\mathrm{eff} = \frac{1}{3}(4 + \frac{1}{P(r)})$ and hence the Newtonian bound of 4/3 is always satisfied for non-negative pressure.

\section{Summary}\label{sec:summary}

In this paper we have investigated the stellar structure of strongly interacting quark stars in the  4D Einstein Gauss-Bonnet theory of gravity for different combinations of the charge $Q$ and 4DEGB coupling constant $\alpha$.  In accord with the lack of a mass gap in the 4DEGB theory \cite{charmousis2022}, we find that even for small $\alpha$ the quark star solutions asymptotically approach the 4DEGB black hole horizon radius, and thus have solutions with smaller radii than the GR Buchdahl/Schwarzschild limits. In general, larger $Q$ and $\alpha$ tend to increase the mass-radius profile of quark stars, with large $\alpha$ suppressing the differences between different charges. These findings are generally consistent with what was found in the regime of weak coupling to the 4DEGB theory \cite{pretel_2022}. For large enough $\alpha$ ($\bar{\alpha} \approx 0.05$), the charged quark star solutions presented in this paper are generally consistent with recent observational data of candidate quark stars HESS J1731-347,
PSR J0030+0451, PSR J0740+6620, and GW190814. Inclusion of an interacting EOS would likely extend this agreement into the lower $\alpha$ regime.

We have found  many additional features in the unexplored regions of parameter space, the most striking of which is that 4DEGB charged quark stars can exist with radii 
not only smaller than the general relativistic Buchdahl bound, but also smaller than the $2M$ Schwarzschild radius.  Such  Extreme Compact Charged Objects represent a possible new state of matter. 
A full analysis of the stability of such objects would be an interesting topic for future study. Furthermore, realistic astrophysical objects tend to have a net angular momentum. While some slowly rotating black hole \cite{gammon_2022} and neutron star \cite{charmousis2022} solutions have been examined in 4DEGB gravity, the case of the rotating quark star has yet to be covered and would be a phenomenologically interesting extension of the current project.

\section*{Acknowledgements}
This work was supported in part by the Natural Sciences and Engineering Research Council of Canada. We would like to thank Dr. Tiberiu Harko for useful correspondence regarding the existing literature.




\bibliographystyle{unsrt}
\cleardoublepage 
\phantomsection  
\renewcommand*{\refname}{References}

\addcontentsline{toc}{chapter}{\textbf{References}}

\bibliography{refs}

\nocite{}




\end{document}